\documentclass[preprint,12pt]{elsarticle}

\usepackage{amssymb}

\usepackage{paralist}
\usepackage[group-separator={,},group-minimum-digits=4]{siunitx}
\usepackage{makecell}
\usepackage{multirow}
\usepackage[flushleft]{threeparttable}
\usepackage{booktabs}
\usepackage{hyperref}
\usepackage{adjustbox}
\usepackage{graphicx}

\journal{}

\begin{document}
\widowpenalty=10000

\begin{frontmatter}



\title{Large scale analysis of gender bias and sexism in song lyrics}


\author[inst1,inst2]{Lorenzo Betti\corref{cor1}}
\cortext[cor1]{Corresponding author: \url{lrn.betti@gmail.com}}

\affiliation[inst1]{organization={ISI Foundation},
            addressline={Via Chisola 5}, 
            city={Turin},
            postcode={10126}, 
            country={Italy}}
            
\affiliation[inst2]{organization={Department of Network and Data Science, Central European University},
            addressline={Quellenstraße 51-55}, 
            city={Vienna},
            postcode={1100}, 
            country={Austria}}

\author[inst3,inst4]{Carlo Abrate}
\author[inst1,inst5]{Andreas Kaltenbrunner}

\affiliation[inst3]{organization={CENTAI},
            addressline={Corso Inghilterra 3}, 
            city={Turin},
            postcode={10138}, 
            country={Italy}}
            
\affiliation[inst4]{organization={Sapienza University},
            addressline={Piazzale Aldo Moro 5}, 
            city={Rome},
            postcode={00185}, 
            country={Italy}}
            
\affiliation[inst5]{organization={Universitat Pompeu Fabra},
            addressline={Tanger 122}, 
            city={Barcelona},
            postcode={08018}, 
            state={Catalonia},
            country={Spain}}
            
\begin{abstract}
We employ Natural Language Processing techniques to analyse \num{377808} English song lyrics from the ``Two Million Song Database" corpus, focusing on the expression of sexism across five decades (1960 - 2010) and the measurement of gender biases.  
Using a sexism classifier, 
we identify sexist lyrics at a larger scale than previous studies using small samples of  manually annotated popular songs. Furthermore, 
we reveal gender biases by measuring associations in word embeddings learned on song lyrics. 
We find sexist content to increase across time, especially from male artists and for popular songs appearing in Billboard charts. Songs are also shown to contain different language biases depending on the gender of the performer, with male solo artist songs containing more and stronger biases. 
This is the first large scale analysis of this type, giving insights into language usage in such an influential part of popular culture. 

\end{abstract}


\end{frontmatter}

\let\thefootnote\relax\footnotetext{Please cite this manuscript as: \\ Betti, L., Abrate, C. \& Kaltenbrunner, A. Large scale analysis of gender bias and sexism in song lyrics. EPJ Data Sci. 12, 10 (2023). \url{https://doi.org/10.1140/epjds/s13688-023-00384-8}}

\section{Introduction}\label{sec:introduction}

Music allows the expression of emotions, feelings and ideas through verbal and auditory language.
Multiple messages can be conveyed by means of this language both at an emotional level, mostly through sounds, and at a verbal level through song lyrics \cite{ransom2015message}.
In particular, musical lyrics can contain verbal messages of various nature, through which ideas of individuals and social groups have been spread in society.
Song lyrics are also a popular expression of culture, and they reflect or maybe even emphasize many social phenomena. They hold ideas that have occurred over time about different social issues, such as gender discrimination and sexism~\cite{cobb2007ambivalent,treat2015influence,adams2006words}.
As stated by Davis in~\cite{davis1985}, song lyrics ``are more than mere mirrors of society; they are a potent force in the shaping of it". 
As such, song lyrics are an important albeit underexplored data source to observe and measure societal changes.

Natural Language Processing (NLP) techniques are an ideal tool for analyzing song lyrics due to their textual nature. These techniques have proven to be effective in improving results for companies, producers, and songwriters in the music industry~\cite{miranda2021quantum}. However, there is growing concern that NLP models may inadvertently learn and perpetuate existing biases encoded in the training data when trained on large text corpora~\cite{hovy2021five}. This raises the possibility of biases being amplified in language models and systems that use them~\cite{abid2021large,shah2020predictive}.

However this undesired property of NLP models can also be turned into an opportunity when it comes to the measurement of linguistic biases present in text corpora. 
Previous studies have demonstrated that word embeddings, which represent words as high-dimensional vectors~\cite{bengio2000neural}, are capable of capturing linguistic biases related to human biases~\cite{caliskan2017semantics}, as well as reflecting stereotypes towards women and ethnic minorities~\cite{bolukbasi2016man,garg2018word}. The magnitude of such biases can vary depending on the domain of the text corpus~\cite{chaloner2019measuring,babaeianjelodar2020quantifying}. These findings provide evidence of the effectiveness of measuring linguistic biases in text through word embeddings.

NLP techniques have also shown promise in identifying hate speech and other forms of discriminatory language in text corpora~\cite{shushkevich2019automatic,jahan2021systematic}. These techniques involve models trained on annotated datasets that have been labeled for various concepts related to hate speech, such as sexist content. The resulting models can then be used to automatically detect and quantify for example the presence of sexist content in new text data.

In this study we take advantage of some of these NLP techniques and analyse English song lyrics regarding two aspects receiving increasing attention: gender-related language bias and sexism in language.
We 
consider two main research questions:
\begin{itemize}
    \item[RQ1:] Is there evidence of sexist content in song lyrics? 
    Can we find differences with respect to artist gender and musical genres across time?
    \item[RQ2:] Are gender and social role stereotypes reflected in song lyrics? Are there differences in the songs with respect to artist gender?
\end{itemize}

To answer these research questions, we use \textbf{Song Lyrics Data} 
from the ``Two Million Song Database" of the WASABI project~\cite{meseguer2017wasabi} to obtain a large corpus of song lyrics and song related information. We enrich the songs' metadata with Billboard chart performance to get the popularity trend of songs that spans from 1960 to 2010.

Then, we apply and adapt the sexism classifier of Samory et al. ~\cite{samory2021call} to explore the presence of \textbf{sexism in song lyrics}. 
This approach allows to analyze the evolution of sexism in song lyrics of the WASABI dataset across time.
We find an increase over time of sexist content in popular song lyrics by male solo artists, and that Hip hop and R\&B and Soul songs have a higher fraction of sexist lyrics when compared to the other analyzed genres.

Finally, we use different word embedding association tests to detect \textbf{(gender) language bias in song lyrics}~\cite{caliskan2017semantics,charlesworth2021gender,bianchi2021sweat}.
Differentiating the analysis by gender, we show the significant presence of language biases in the WASABI dataset, in particular for male solo artists, in contrast to more gender neutral bias of female solo artists.

The main contributions of this work consist in providing (to our best knowledge):
\begin{itemize}
    \item the first large scale and longitudinal exploratory data analysis employing an automatic method to identify song lyrics containing sexist content,
    \item the first extensive study of language bias in song lyrics segregating by artist gender.
\end{itemize}

\section{Related work}
The analysis of gender stereotypes and sexism expressed in language with NLP methods has been a growing area of research in recent years \cite{stanczak2021survey}. These methods are closely entangled with the detection and mitigation (i.e. removal or reduction) of gender bias in NLP models and their output (see for example Sun et al.~\cite{sun2019mitigating} for a review on mitigation).
A promising approach for both, detection and mitigation, relies on measuring gender bias from the association between word vectors, for example through their cosine similarity.
Bolukbasi et al.~\cite{bolukbasi2016man} proposed to first identify a gender direction 
defined by the subspace spanned by gendered words (e.g., she and he, women and man). Then, they quantify gender bias as the projection of ideally gender-neutral words (e.g., job and profession names) onto the gender subspace. 

This idea, combined with averaging methods and hypothesis testing has later led to the development of the 
Word Embedding Association Test (WEAT)~\cite{caliskan2017semantics}. The significance and sensitivity of WEAT  has been analyzed by 
\cite{ethayarajh-etal-2019-understanding} and it has been successfully employed for example in \cite{charlesworth2021gender} to quantify gender
stereotypes in language corpora, adding a single category version (SC-WEAT) to the analysis.
WEAT has later been extended further to SWEAT~\cite{bianchi2021sweat} to compare the relative polarization of two corpora (like male and female authored lyrics in our case).  
We will exploit WEAT, SC-WEAT and SWEAT here and describe them in more detail in Section~\ref{sec:methods_estimating_language_biases}.
Besides word embeddings, similar approaches have been used to measure gender bias in large language models~\cite{babaeianjelodar2020quantifying,nadeem2021stereoset}.

Part of our work furthermore builds on the results of a sexism classifier developed by Samory et al.~\cite{samory2021call}. Since the exact definition of misogyny and sexism may be under discussion \cite{manne2017down}, the authors of \cite{samory2021call}  took into account different dimensions of sexism to increase model validity and furthermore improved the model reliability through including adversarial examples.
Other noteworthy approaches to sexism detection use support vector machines, sequence-to-sequence models and a FastText classifier~\cite{jha2017does}; a BERT-based architecture to detect misogyny and aggression simultaneously on social media~\cite{samghabadi2020aggression} or compare different models for identifying misogyny across languages and domains on Twitter data \cite{pamungkas2020misogyny}.
We refer to the following reviews for a more comprehensive overview about automatic misogyny and hate speech detection~\cite{shushkevich2019automatic,jahan2021systematic}.

In relation to our research questions, most of the works on gender stereotypes and sexism in song lyrics have targeted popular songs using manual content analysis, which allows studying fine-grained constructs related to sexism such as objectification and sexualization~\cite{Madanikia2014themes,hall2012sexualization}.
These studies analysed popular songs from the 1960s to the 2000s showing that sexualization and mentions of sexual desire increased dramatically only after the 1990s, while mentions of love decreased. This is interpreted as a signal of the increase of sexism in songs because lust in the absence of love is likely to objectify the object of the desire~\cite{Madanikia2014themes}. Other works consider both the differences between male and female artists and the gender of the target being objectified. In~\cite{smiler2017want}, authors analyze the same time period considered in our work, finding female artists more likely to sing about love, while men are more likely to objectify others (both men and women), with a stronger emphasis towards women. Flynn et al.~\cite{flynn2016objectification} confirmed these findings, adding that women were more likely to objectify themselves than men do. 
Moreover, Rap, Hip hop and R\&B songs are the genres with the most objectification.
These works rely on different aspects of the concept of sexism, which is possible to code through manual content analysis. Although our methodology can not distinguish between such nuances, it allows to analyse song lyrics at a larger scale.

Here we are more interested in studies using large datasets like the one from~\cite{barman2019decoding}, which analyses bias in half a million song lyrics using WEAT scores~\cite{caliskan2017semantics}. This study does not segregate its results neither by gender nor in the temporal dimension and finds that bias in songs is strongest in relation to gender stereotypes and career paths. However, also gender biases in relation to Math vs Arts and Science vs Arts are found, meaning that both math and science words are more closely related to male terms while females terms are more closely related to art. All these biases are similar (albeit a bit smaller) to what can be observed in a large internet crawl of texts~\cite{caliskan2017semantics}, thus indicating that biases present in song lyrics mirrors the biases that exist in society. Recently, in a study performed in parallel to ours it was found that in song lyrics men are more likely than women to be associated to traits depicting them as competent, even though this bias becomes weaker going forward in time~\cite{Boghrati2022Quantifying}. The authors of this work, albeit incorporating a temporal dimension, used a simpler not standardized metric of gender bias and measured only a single trait, whereas our analysis investigates several traits and uses different association tests used as well in other published works~\cite{caliskan2017semantics, charlesworth2021gender, bianchi2021sweat}. 

Other tasks for which song lyrics have been used for is song mood \cite{hu2009lyric} and sentiment \cite{xia2008sentiment} classification, or together with audio features for genre classification \cite{mayer2011musical} and song popularity \cite{martin2020multimodal}. 
The performance in the later two tasks with lyrics alone has been explored in \cite{barman2019s}.

\section{Data Collection and Methods}\label{sec:methods}

\subsection*{Data collection and filtering}\label{sec:results_data_collection}

The WASABI database is a knowledge base that includes data about 77K artists, 200K albums, and more than two million songs~\cite{meseguer2017wasabi}. We queried the database for all solo artists and groups having published more than 10 songs, and collected all their English\footnote{The WASABI database provides the language of song lyrics through the ``language\_detect" field.} song lyrics published between 1960 and 2009. The database contains information about solo artists and band members, including their gender. We assigned the gender label ``male" (``female") to bands composed of only male (female) members, and  ``mixed" to bands with both male and female members. We discarded solo artists without gender information, as well as groups with at least one member without it. The database does not include other gender identities, except for 9 artists whose gender is labeled as ``Other'' and accounting for a total of 44 songs. We thus decided to consider artists' gender as binary, although we acknowledge this limitation that prevents us from taking into account other gender identities. Furthermore, we filtered out all the songs whose lyrics or publication year are unavailable, whose lyrics are shorter than 10 words or composed of less than 4 lines. After removing duplicate lyrics (details in~\ref{appendix:duplicate_detection}), our final dataset consists of \num{377808} song lyrics: \num{244146} are performed by \num{7131} solo artists and the remaining \num{133662} by \num{4294} groups. Finally, we retrieved also the song genres, which was not available for \num{46482} songs (12.3\%).  The number of unique genres was reduced by replacing them with their corresponding top-level genres\footnote{We used the taxonomy of the Wikipedia page of popular genres~\url{https://en.wikipedia.org/wiki/List\_of\_music\_genres\_and\_styles}.}.

To identify popular songs in the WASABI dataset, 
we retrieved the Billboard Hot 100 weekly charts, composed of all-genre weekly song charts released since 1958~\cite{BillDataWebsite}. We furthermore extracted the top 10 songs from all the charts. 
We were able to map \num{10798} out of \num{24180} unique songs (44.7\%) of the Billboard, and \num{2608} out of \num{4348} unique songs (60.0\%) of the Billboard top 10 charts by using approximate string matching to match both the title and the artist of songs (details in~\ref{appendix:matching_datasets}).
In the following, when referring to songs in Billboard or in Billboard top 10 charts, we refer to these sets of songs that we mapped to the main WASABI dataset. 

\subsection{Sexism detection}\label{sec:methods_sexism_detection}

We fine tuned a BERT\footnote{\url{https://tfhub.dev/google/bert\_uncased\_L-12\_H-768\_A-12/1}} classifier to detect sexist passages in texts using the dataset and the code provided by Samory et al.~\cite{samory2021call}. The dataset contains texts manually labeled following a scale-based codebook that operationalizes the concept of sexism. The codebook consists of four non-overlapping categories resulting from the review of psychological scales measuring sexism and related constructs. These include a broad range of sexist content that can be present in song lyrics such as behavioral expectations, stereotypes \& comparisons, endorsements of inequality, and denying inequality \& rejection of feminism. In addition to sexist content, the codebook also includes categories that take into account sexist phrasing, in order to distinguish sexist content from texts containing only uncivil content or common profanity.
This makes the classifier perform well on out-of-domain data, also thanks to the inclusion of human-written adversarial examples. We refer to Samory et al.~\cite{samory2021call} for further details on the codebook.

Since the original dataset consists of short texts, we adapted the representation of the input to classify song lyrics. In detail, lyrics are divided into batches composed of groups of four lines, each of them sharing two lines with the previous and following group. A batch is considered to contain sexist content if the model outputs a probability higher than a certain threshold. Whenever a song lyric contains at least one batch identified as sexist, we propagate the sexist label to the song. To verify the performance on song lyrics, we evaluated the classifier on an external dataset~\cite{sexist_dataset} of 190 lyrics\footnote{Mapped to WASABI using the same approach used for the Billboard charts, described in~\ref{appendix:matching_datasets}.}, 40.5\%  of which are considered to contain sexist content. For the optimal classification threshold of $0.725$, the classifier achieves a precision of 0.73. We discuss in~\ref{appendix:scores_sexism_classif} the results obtained for different classification thresholds and requiring a larger number of batches labeled as sexist to propagate the label to the whole song, showing that our main findings are not affected by these choices.

\subsection{Language biases from word embeddings}\label{sec:methods_estimating_language_biases}

Words can be represented by vectors by leveraging the co-occurrence of nearby words in such a way that vectors that are close to each other represent words sharing a similar semantic meaning~\cite{mikolov2013efficient}. This representation of words is able to encode word analogies (e.g., ``King" is to ``Man" as ``Queen" is to ``Woman"), but it may also encode language biases and stereotypical analogies that are present in the training corpus~\cite{caliskan2017semantics}.
For example, from a higher co-occurrence of pleasant words next to music instrument than to weapon words, we can likely expect word embeddings of unpleasant words to be closer to weapon than musical instrument words.
We measured language biases in three different subsets of the lyrics corpus: all lyrics of solo artists, lyrics of male solo artists, and lyrics of female solo artists. 
This allowed us to compare the biases that are present in song lyrics performed by male and female artists.
In the following, we list the methodologies we used to measure these kinds of associations in word embeddings.

{\bf WEAT:} 
The Word Embedding Association Test (WEAT) measures the association between two sets of target words $X$ and $Y$ (e.g., music instruments and weapons), and two sets of attribute words $A$ and $B$ (e.g., pleasant and unpleasant words). The WEAT tests the null hypothesis that there is no difference between the two sets of target words in terms of their relative similarity to the two sets of attribute words~\cite{caliskan2017semantics}. 
The similarity between two words $a$ and $b$ is defined as the cosine similarity $cos(\vec{a}, \vec{b})$ between their representation in a vector space. 
The effect size is defined as:
$$ \frac{\text{mean}_{x\in X} s\left( x, A, B\right) - \text{mean}_{y\in Y} s\left( y, A, B\right)}{\text{pooled\_std\_dev}_{w\in X \cup Y} s\left( w, A, B\right)} $$
where
\begin{equation}\label{eq:s}
    s\left( w, A, B\right) = \text{mean}_{a \in A} \cos\left(\vec{w}, \vec{a}\right) - \text{mean}_{b \in B} \cos(\vec{w}, \vec{b}\,)
\end{equation}
and $\text{pooled\_std\_dev}$ is the pooled standard deviation~\cite{cohen2013statistical}. \footnote{We found different implementations of the WEAT, which compute the effect size using either the standard deviation or the pooled standard deviation (i.e. a weighted average of standard deviations). We opted for the latter solution because it reflects the definition of the effect size in terms of Cohen's D defined in Caliskan et al.~\cite{caliskan2017semantics}.} 
From this definition, positive values of the effect size indicate that the words in $X$ are more similar to the words in $A$ than in $B$, and the words in $Y$ are more similar to the words in $B$ than in $A$. Recalling the above example, we would expect a positive effect size.  
The significance is computed in the following way.
The test statistic is defined as:
$$ S\left(X,Y,A,B \right) = \sum\limits_{x\in X} s\left( x, A, B\right) - \sum\limits_{y\in Y} s\left( y, A, B\right)$$
Given \num{1000} random partitions of equal size for the union of the two target sets $\left\{\left(X_i, Y_i \right) \right\}_{i=1,\dots,1000}$, the one-sided P value is computed as $$ Pr_i\left[S\left(X_i,Y_i,A,B \right)>S\left(X,Y,A,B \right) \right].$$

{\bf SC-WEAT:}
Since the WEAT condensates into one effect size the relative associations of attribute sets against two target sets, the similarity against one single target set is lost.
We used the Single Category Word Embedding Association Test (SC-WEAT) to disentangle this relative association~\cite{charlesworth2021gender}. 
The SC-WEAT measures the association between one set of target words $W$ and two sets of attribute words $A$ and $B$.
If the SC-WEAT score is positive (negative), words in $W$ are more (less) similar to the words in $A$ compared to $B$.
The procedure to compute the significance is analogous to the one described for the WEAT.
Note that the WEAT and the SC-WEAT are not redundant because the former measures the relative association of $X$ and $Y$ with $A$ and $B$, whereas the SC-WEAT evaluates the association of one target set per time, and their combination can add information to the language biases analysis. 

{\bf SWEAT:}
The methods described above measure associations between word sets whose embeddings are learnt from a single text corpus. To compare the association between two corpora, we employed the Sliced Word Embedding Association Test (SWEAT)~\cite{bianchi2021sweat}. 
Given the word vectors learnt on two corpora $\mathcal{D}_1$ and $\mathcal{D}_2$, the SWEAT score is defined as:
$$ S\left( W, A, B, \mathcal{D}_1, \mathcal{D}_2\right) = \sum\limits_{w\in W} s\left(w, A, B, \mathcal{D}_1 \right) - \sum\limits_{w\in W} s\left(w, A, B, \mathcal{D}_2 \right) $$
where $ s\left(w, A, B, \mathcal{D} \right) $ has the same form as in Equation~\ref{eq:s}, but word vectors are taken from the distributional representation $\mathcal{D}$. 
A positive SWEAT score indicates that the word vectors of a target set in $\mathcal{D}_1$ are relatively more associated to the words in $A$ than in $B$, and the word vectors in $\mathcal{D}_2$ are relatively more associated to the words in $B$ than in $A$.
Again, significance is computed in the same way as for the WEAT.
Here, $\mathcal{D}_1$ and $\mathcal{D}_2$ refer to the male and female corpus respectively. 

{\bf Selection of attribute and target words:} Table~\ref{tab:words_used} in the Appendix reports the sets of target and attribute words used, borrowed from~\cite{caliskan2017semantics} and~\cite{chaloner2019measuring}. We slightly modified the word sets to account for rare or missing words. For this, we first removed words occurring less than five times in at least one of the three corpora. Then, whenever pairs of attribute (target) sets contain a different number of words, we removed the least frequent words from the larger set until the two sets have the same size. In doing this, we defined the frequency of a word as the minimum of its frequencies in the three corpora under analysis. For the list of male and female proper names, we used the most frequent proper names in the lyrics corpora.

{\bf Learning word vectors:}
We used the Gensim implementation of Word2Vec~\cite{rehurek2011gensim} to learn word embeddings from scratch separately for the three corpora of song lyrics. 
To account for the potential variability that arises from different initializations of the word vectors, we conducted the association tests on word embeddings obtained from five independent runs of the Word2Vec algorithm.
We then report the score obtained from the first iteration and deem the result to be statistically significant only if all five iterations yielded a significant score at a certain level.
In other words, we conducted the association tests five times using different initializations to ensure the robustness of our results.

\section{Results}
\label{sec:results}

\subsection{Basic statistics of the dataset}\label{sec:basic_stats}

\begin{figure}[!tb]
    \centering
    \includegraphics[width=0.96\columnwidth]{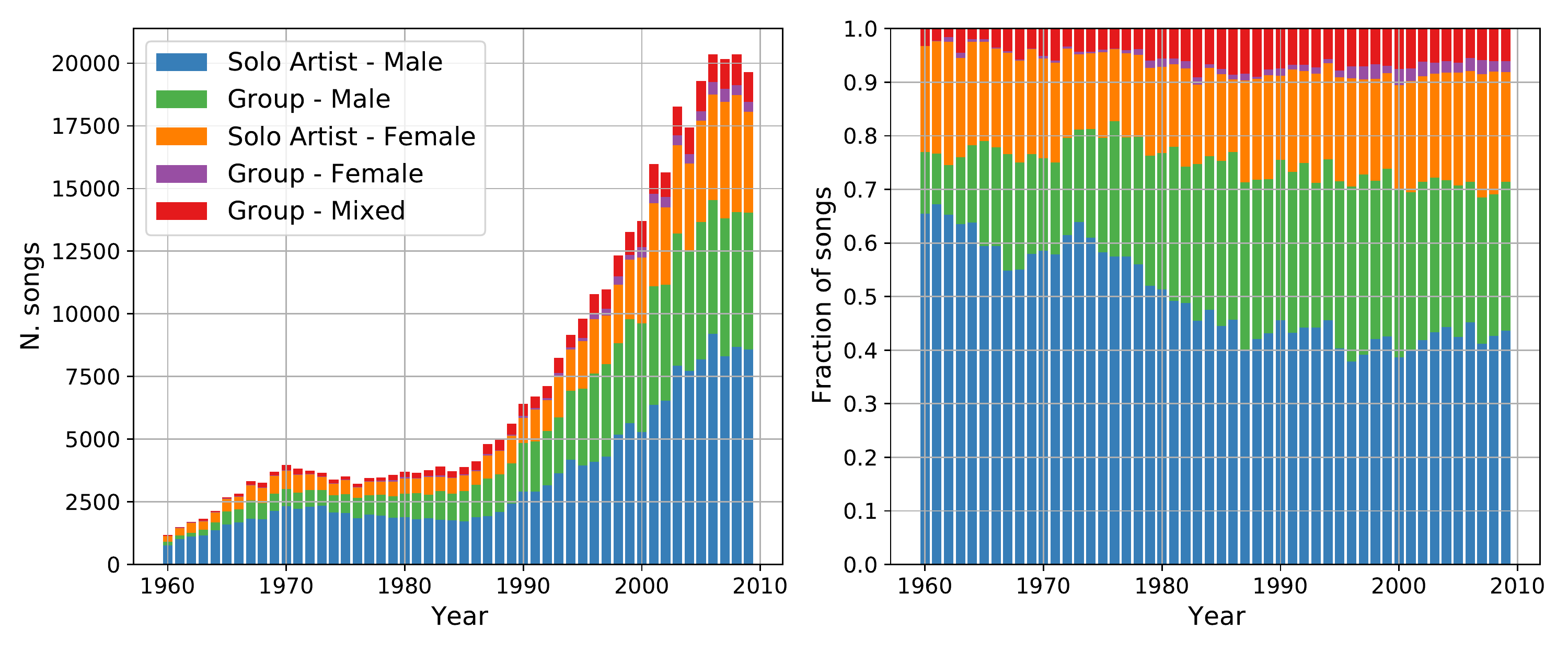}
    \caption{Yearly number of songs (left) and relative fraction of songs (right) of the WASABI dataset. Colors refer to different artist type and gender.}
    \label{fig:n_songs_across_time}
\end{figure}

Table~\ref{tab:basic_stats_dataset} reports the number of songs for each combination of artist type and gender. Solo artists are more represented by males than females, with more than the double of the number of songs. This unbalance is exacerbated for groups, where female and mixed groups account for 1.7\% and 6.2\% of the songs. This observation holds for the songs in Billboard and Billboard top 10 charts as well, where the gender inequality in English song charts is well documented in the literature~\cite{Lanfrance@gender, anglada2021popular}.  
The distribution of publication years of the songs in our dataset is shown in the left plot of Figure~\ref{fig:n_songs_across_time}. The total number of songs per year is not uniformly distributed across time. There are about \num{3000} songs per year from the 1970s until the 1990s with a subsequent steady increase reaching around \num{20000} songs per year in the late 2000s. However, the relative fraction of songs performed by male and female artists remains relatively constant, as can be seen in the right plot of Figure~\ref{fig:n_songs_across_time}. Indeed, male artists (in groups or as solo artist) contribute with a fraction of songs between 70\% and 80\% over the whole time span. 

\begin{table}[t!]
\caption{Basic statistics of the dataset. }
\label{tab:basic_stats_dataset}
\centering
    \begin{tabular}{@{}l|r|r|r|r|r@{}} 
        \makecell{Artist\\Type} & Gender & Artists & Songs & Billboard & \makecell{Billboard \\ (top 10)}  \\ \hline
        \hline
        \multirow{2}*{\makecell{Solo\\Artist}} & Male & \makecell[r]{\num{4770}\\41.8\%} & \makecell[r]{\num{171324}\\45.3\%} & \makecell[r]{\num{5071}\\47.0\%} & \makecell[r]{\num{1103}\\42.3\%}  \\ \cline{2-6}
        & Female & \makecell[r]{\num{2361}\\20.7\%} & \makecell[r]{\num{72822}\\19.3\%} & \makecell[r]{\num{2412}\\22.3\%} & \makecell[r]{\num{664}\\25.5\%}  \\ \hline
        \multirow{3}*{Group} & Male & \makecell[r]{\num{3281}\\28.7\%} & \makecell[r]{\num{103729}\\27.5\%} & \makecell[r]{\num{2493}\\23.1\%} & \makecell[r]{\num{629}\\24.1\%}  \\ \cline{2-6}
        & Female & \makecell[r]{\num{306}\\ 2.7\%} & \makecell[r]{\num{6416}\\1.7\%} & \makecell[r]{\num{212}\\2.0\%} & \makecell[r]{\num{56}\\2.1\%}  \\ \cline{2-6}
        & Mixed & \makecell[r]{\num{707} \\ 6.2\%}& \makecell[r]{\num{23517}\\6.2\%} & \makecell[r]{\num{610}\\5.6\%} & \makecell[r]{\num{156}\\6.0\%}  \\
        \hline
        \multicolumn{2}{l|}{Total} &\num{11425} & \num{377808} & \num{10798} & \num{2608}\\
    \end{tabular}
\end{table}

\begin{figure*}[!tbh]
    \centering
    \includegraphics[width=.95\textwidth]{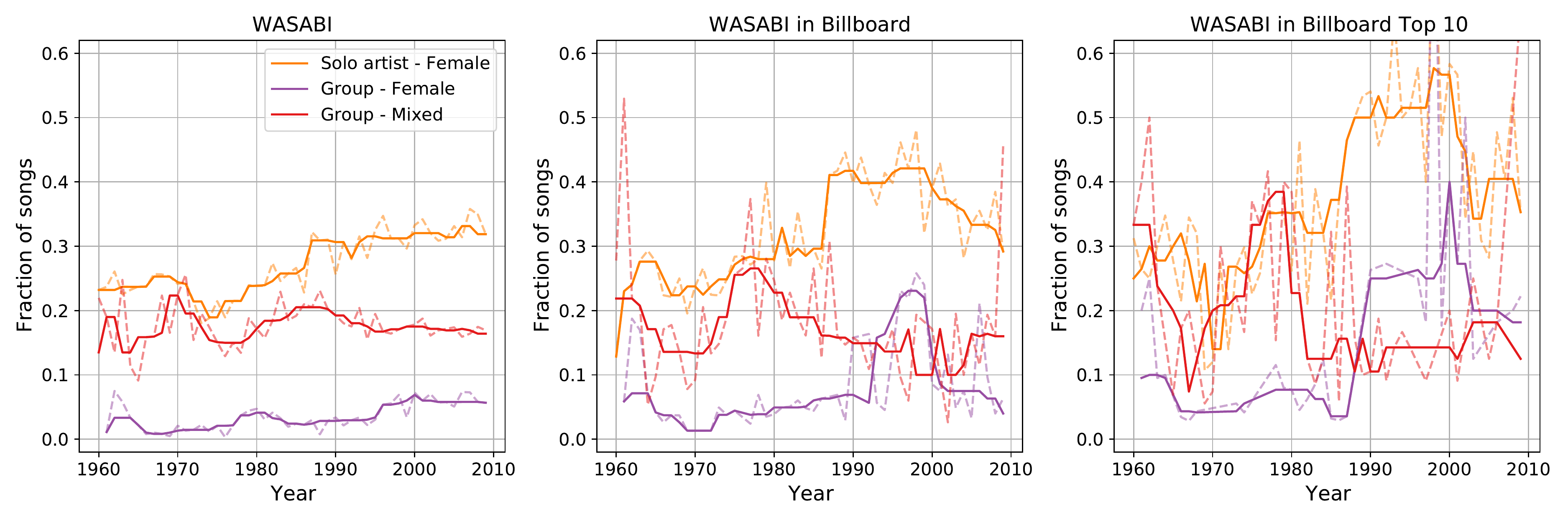}
    \caption{Yearly fraction of songs by female solo artists, female groups, and mixed groups. The three plots refer to the (filtered) WASABI dataset (left), songs in Billboard charts (center), and songs reaching Billboard top 10 (right).
    Dashed lines are raw fractions of songs, solid lines a median filter with window $=5$ years. Two data points in Billboard top 10 are out of the figure's scale.
    }
    \label{fig:frac_fem_songs_across_datasets_time}
\end{figure*}

Figure~\ref{fig:frac_fem_songs_across_datasets_time} shows the increase of the fraction of songs by female solo artists across time in all the subsets of the dataset. In particular, the Billboard and Billboard top 10 charts show a sharp increase between 1980 and 1990 where the fraction of female artist songs increases by around 10\%, reaching 40\% and 50\% respectively. Then, during the 2000s this fraction decreased again to values of the mid-80s. We refer to~\ref{appendix:descriptive_stats} for the fraction of songs  segregated by genre across time.

\subsection{Sexist song lyrics}\label{sec:sexism_classif_results}

Table~\ref{tab:basic_stats_dataset_sexism_optimal_thresholdd} shows the proportion of songs containing sexist passages identified for each artist type and gender. The classifier identifies \num{89462} (23.7\%) lyrics in the WASABI dataset to contain sexist passages. Artists and groups have different fractions of sexist lyrics: 30\% of male solo artist songs are classified as sexist,
compared to 16\% to 20\% of songs of the other groups of artists.
The fraction of sexist songs in Billboard and Billboard top 10 charts is at least 10\% higher than the one of the whole WASABI dataset. 
This observation suggests that popular songs are more likely to contain sexist content than an average song, even more so if they reach a top position in the charts. This finding is consistent for different classifications thresholds (details in~\ref{appendix:scores_sexism_classif}). 

\begin{table}[t!]
\caption{Percentage of songs containing sexist passages identified for each artist type and gender. Percentages correspond to the fraction of sexist lyrics within the artist type and gender.}
\label{tab:basic_stats_dataset_sexism_optimal_thresholdd}
\centering
    \begin{tabular}{@{}l|l||p{1.8cm}|p{1.8cm}|p{1.8cm}} 
       \makecell[c]{Artist\\Type} & \makecell[c]{Gender} & \makecell[c]{Songs} & \makecell[c]{Billboard} & \makecell[c]{Billboard\\(top 10)} \\
        \hline
        \hline
        \multirow{2}*{\makecell{Solo\\Artist}} & \makecell[l]{Male}  &  \makecell[r]{30.0\%} &  \makecell[r]{43.4\%} &  \makecell[r]{48.4\%}  \\ \cline{2-5}
        & \makecell[l]{Female}  &\makecell[r]{19.6\%} & \makecell[r]{29.8\%} & \makecell[r]{30.9\%} \\ \hline
        \multirow{3}*{Group} & \makecell[l]{Male \\only}  & \makecell[r]{18.0\%} &  \makecell[r]{32.0\%} &  \makecell[r]{33.2\%} \\ \cline{2-5}
        & \makecell[l]{Female \\only}  & \makecell[r]{20.0\%} &  \makecell[r]{34.4\%} &  \makecell[r]{42.9\%} \\ \cline{2-5}
        & \makecell{Mixed}  &  \makecell[r]{16.2\%} &  \makecell[r]{26.4\%} &  \makecell[r]{30.1\%}\\ 
       
    \end{tabular}
\end{table}

\begin{figure}[!t]
    \centering
    \includegraphics[width=.95\textwidth]{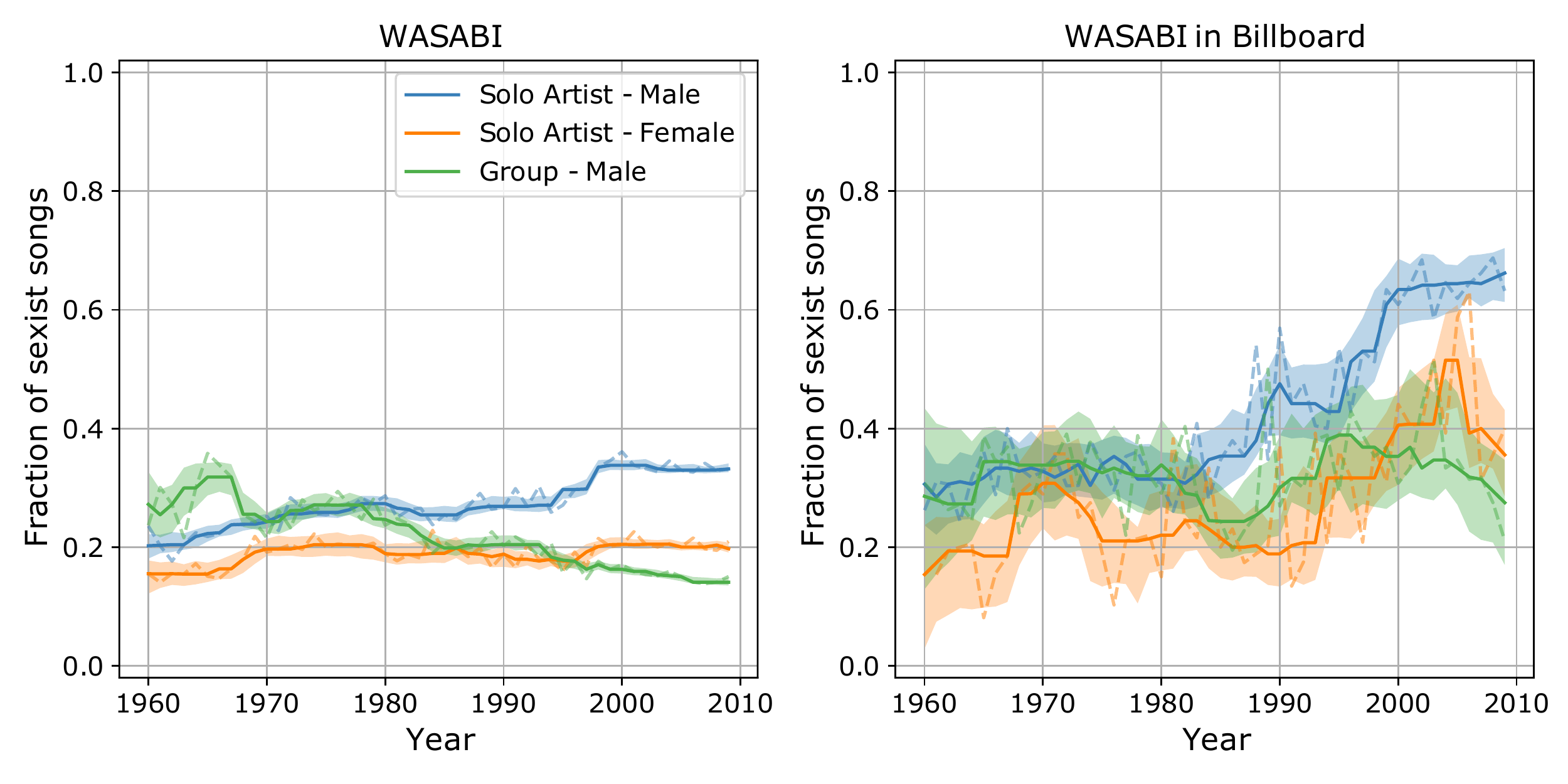}
    \caption{
    Fraction of songs containing sexist content. Colors refer to different artist type and gender.
    Dashed lines are the raw fractions of songs, and solid lines with 95\% confidence intervals were obtained using a median filter with a window equal to 5 years. }
    \label{fig:frac_sexist_songs_across_datasets_time}
\end{figure}

Furthermore, a larger fraction of male solo artist songs contains sexist content when compared to those of female solo artists. 
We observe that male and female artists display different trends for what concerns their percentage of songs classified as sexist over time, but male solo artists consistently publish relatively more sexist songs in the whole time span. Figure~\ref{fig:frac_sexist_songs_across_datasets_time} shows the fraction of songs containing sexist passages for each artist type and gender. The share of female solo artist songs in the WASABI dataset with sexist lyrics remains relatively constant over time at around 20\%, but, even though the share of male solo artist sexist songs becomes larger going forward in time, this trend is even stronger for popular songs in which we can identify a sharp increase starting around the mid-80s. At the end of the time span under analysis, more than 60\% of male solo artist songs on Billboard are found to contain sexist passages.
The share of female solo artist songs on Billboard with sexist lyrics increases over time as well, but during the 2000s this share is 20\% lower when compared to male solo artists. Differently, group artists, regardless of members' gender, do not display a relevant increase of sexist songs. Figure~\ref{fig:frac_sexist_songs_across_datasets_time} shows only the group male artists to not overload the figure, but the female and mixed group artists display a similar trend. Interestingly, male solo artists perform relatively more sexist songs than male group artists.
This might signal a difference between male solo artists and groups in terms of subjects and themes covered in their songs.
As observed previously, the figure shows also that Billboard charts, compared to the WASABI dataset, attracted more songs with sexist content starting from the 1990s.

After this aggregate analysis we now inspect how different genres contribute to the observed trends.\footnote{Note that the genre information is missing in 382 (3.5\%) and 65 (2.5\%) songs in Billboard and Billboard top 10 charts respectively.}
In particular, we focus on the four most popular genres in our data-sets: Pop, Rock, Hip hop, R\&B and soul. The changes of their frequency over time in the data set is depicted in Figure~\ref{fig:frac_songs_across_datasets_time_genre} in the Appendix, while Figure~\ref{fig:frac_sexist_songs_across_datasets_time_genre} shows the fraction of songs containing sexist passages of male and female solo artists as well as male only groups for Pop, Rock, Hip hop, and R\&B and soul\footnote{Again we omitted female and mixed groups from this analysis.
}. 
A majority of the Hip hop songs are classified as sexist, regardless of artists' gender. 
Pop and Rock songs do not show a substantial difference between male and female solo artists, but male artists perform relatively more sexist songs than female artists for the Pop genre. In both cases, these fractions are lower than the Hip hop genre. Lastly, R\&B and soul songs display a steady increase of sexist lyrics over time, with male solo artists having a higher fraction of sexist songs than female artists. 

\begin{figure}[hbt!]
    \centering
    \includegraphics[width=\textwidth,height=.88\textheight,keepaspectratio]{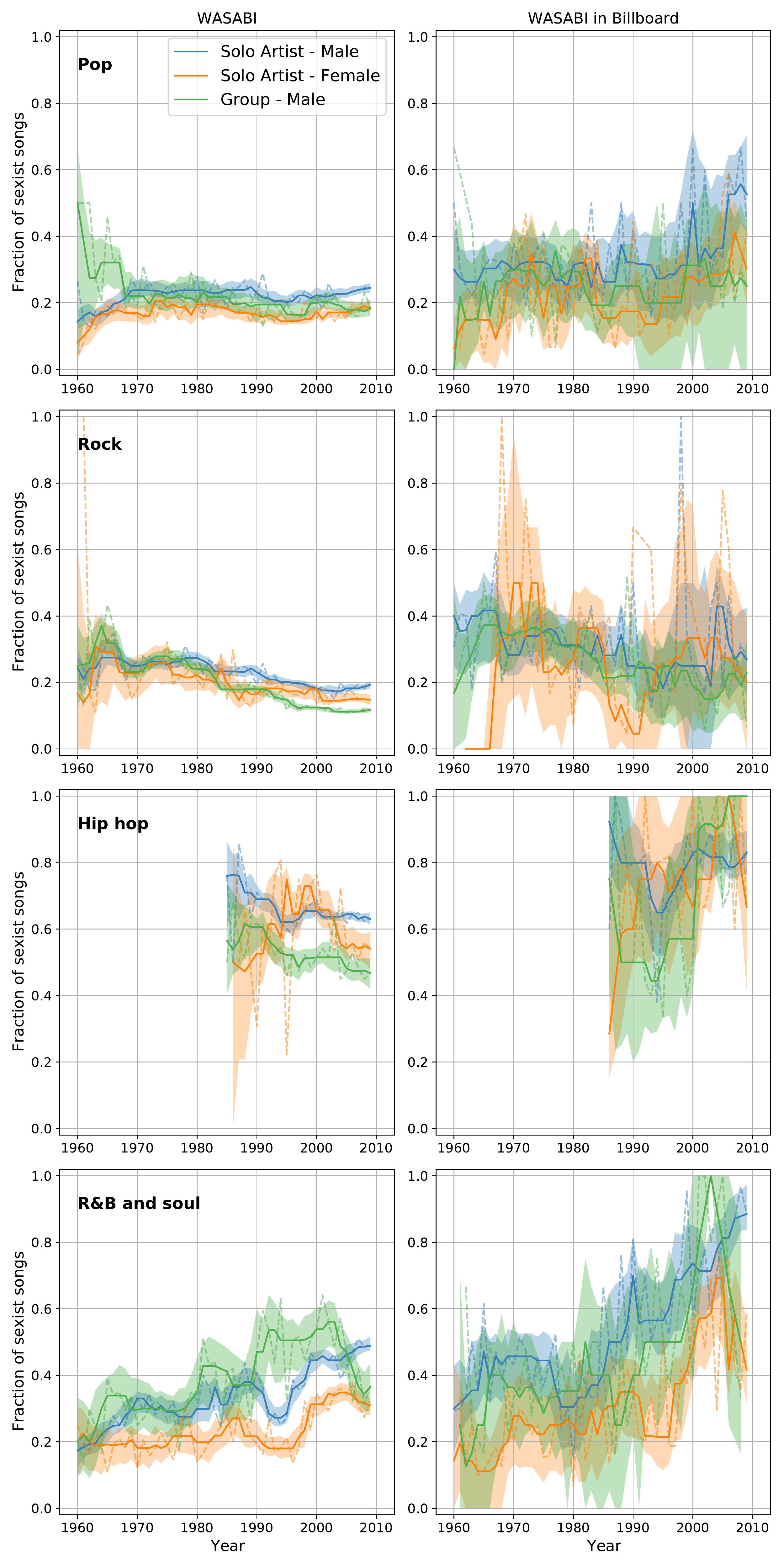}
    \caption{ 
    Fraction of songs containing sexist content. Each row refers to a genre and columns to two subsets of the WASABI dataset. 
    Details as in Figure~\ref{fig:frac_sexist_songs_across_datasets_time}.
    }
    \label{fig:frac_sexist_songs_across_datasets_time_genre}
\end{figure}

\subsection{Language bias in song lyrics}\label{sec:bias_lyrics}

Here we investigate the corpora of song lyrics for potential language bias, showing the 
results of the different word embedding association tests defined in Section~\ref{sec:methods_estimating_language_biases}: SC-WEAT (measures the difference between two sets of attribute words and a target set), WEAT (measures the relative similarity of the two target sets with two sets of attribute words), and SWEAT (compares the differences for a target set in two corpora, i.e. male and female corpora).

The corresponding results are shown in Table~\ref{tab:results_weat} with positive WEAT effect sizes (last column) for all the tests, indicating  that words in the target set $X$ (e.g., male names)  are more similar to words in $A$ (e.g., career) than $B$ (e.g., family), and words in the target set $Y$ (e.g., female names)  are more similar to words in $B$ than $A$ (e.g., family vs career).  However, the magnitudes and significance levels of these paired associations are different. We now explain the detailed results grouped by the tested sets of words.

{\bf Pleasant vs. Unpleasant words:} These associations are considered to be universally accepted stereotypes for humans~\cite{guo2021detecting}: pleasant words to flowers and musical instruments, and unpleasant words to insects and weapons.
Experiments involving human subjects also found these associations~\cite{Greenwald1998-lr}. We thus expect to find the same in song lyrics as well, regardless of the gender of the artist. 
Indeed, we found all these associations to be statistically significant for both pairs of target sets in all the three corpora (positive and statistically significant WEAT effect size). 
The fact that the trained word vectors capture these trivial stereotypes makes us confident that the word vectors have learned meaningful associations (see~\ref{appendix:word_sets} for a more detailed discussion of these results).
Besides that, we observe a negative and statistically significant SWEAT score for Musical instruments, meaning that Musical instrument words are closer to Unpleasant than Pleasant words in the male corpus while being closer to Pleasant than Unpleasant words in the female corpus.
This might hint at a deeper difference in the way male and female artists refer to music or the tools (instruments) they use to produce it.

{\bf Career vs. Family:} 
We find that Career words are significantly closer to Male than to Female proper names in the male lyrics (SC-WEAT X $=2.76$), as well as that Female names are significantly closer to Family than to Career words (SC-WEAT Y $=-1.04$).
This is not significant if we consider nouns and pronouns instead of names and may be explained by song lyrics being more gender-stereotypical when songs address a specific person mentioned by their proper name.

Interestingly, we do not observe the same bias in female lyrics,
where the SC-WEAT score for Female names is close to 0 and we therefore observe a significant difference between the two corpora (SWEAT $=-0.92$ for Female names). Furthermore, the SC-WEAT score for Career is much smaller ($0.45$), but this does not translate into a significant numerical difference between male and female corpora.
However, a significant difference is also found for Male names (SWEAT $=-0.68$). Although the corresponding SC-WEAT scores are not significant, we observe a positive value ($0.94$) in female lyrics and a negative score in male lyrics ($-0.24$). In other words, Male names are closer to Career than to Family terms in female artist lyrics than in those from male artists.

We observe thus that male solo artists associate Female names closer with Family and female solo artists Male names closer to Career, while the opposite is not the case. There are no biases in female lyrics of female names towards Family terms and in male lyrics of Male names towards Career terms (if anything, it would be in the opposite direction, i.e. a negative SC-WEAT score of $-0.24$). However, when considering career words alone there is a bias in the male corpus of Career terms towards Male names (SC-WEAT X $=2.76$).  So Career terms are more likely to be mentioned together with Male names, while if Male names are mentioned there, it is slightly more likely to be in relation to Family words.

{\bf MatSci vs. Arts:} When analysing a potential gender bias of Mathematics or Science vs. Arts terms we observe that female lyrics show a positive and significant WEAT score ($1.29$), and a negative and significant single category association for Arts, indicating a higher association of Arts words to Female than to Male nouns and pronouns (SC-WEAT Y $= -1.26$). 
None of the other tests is statistically significant for the male and all corpora, nor the comparison of the male and female corpora through SWEAT. 

{\bf Intelligence vs. Appearance and Strength vs. Weakness:} Male solo artists associate Strength words significantly more to Male than Female nouns and pronouns (WEAT $= 1.02$) with a positive and significant single category association for Strength words (SC-WEAT X $= 1.24$). 
Although we also find a similar significant relation for female artists lyrics (SC-WEAT X $= 0.70$), the direct comparison with the male corpus gives a positive and significant SWEAT score for Strength (0.32) terms. 
This indicates the association of Strength words to Male terms is stronger in lyrics of male than female solo artists.

\begin{table}[!t]
\caption{Results of the association tests performed on three different lyrics corpora: male solo artists, female solo artists and their union (all). Target sets X or Y resulting in statistically significant SWEAT scores (comparing male with the female corpus) highlighted in bold with their corresponding score
(* $p<0.10$, ** $p<0.05$).
}

\label{tab:results_weat}
\centering
\resizebox{\textwidth}{!}{
\begin{tabular}{@{}l@{\hskip 3mm}lll@{\hskip 3mm}lrrr@{}}

\cline{1-8}
A     & B      &    X         &     Y  &   corpus   & \makecell[b]{SC-WEAT\\X}  & \makecell[b]{SC-WEAT\\Y}  & \makecell{WEAT\\Effect\\size}  \\
\cline{1-8}
\multirow{6}{*}{Pleasant} & \multirow{6}{*}{Unpleasant} & \multirow{3}{*}{Flowers} & \multirow{3}{*}{Insects} & all &   1.34** &  -0.70** &      1.73** \\
         &            &                     &         & female &   1.31** &  -0.75** &      2.00** \\
         &            &                     &         & male &   0.90** &  -0.78** &      1.59** \\
\cline{3-8}
\cline{4-8}
         &            & \multirow{3}{*}{\bfseries\makecell[l]{Musical\\instruments*\\-0.24}} & \multirow{3}{*}{Weapons} & all &     0.41\hphantom{**} &  -2.79** &      2.97** \\
         &            &                     &         & female &   0.61** &  -1.27** &      1.86** \\
         &            &                     &         & male &     0.23\hphantom{**} &  -1.82** &      1.99** \\
\cline{1-8}
\cline{2-8}
\multirow{3}{*}{Career} & \multirow{3}{*}{Family} & %
         \multirow{3}{*}{\bfseries\makecell[l]{Male\\ names**\\-0.68}} & \multirow{3}{*}{\bfseries\makecell[l]{Female\\ names**\\-0.92}} & all &    -0.05\hphantom{**} &  -0.76** &      0.89** \\
         &            &                     &         & female &     0.94\hphantom{**} &    -0.00\hphantom{**} &        0.57\hphantom{**} \\
         &            &                     &         & male &    -0.24\hphantom{**} &  -1.04** &      0.84** \\
\cline{1-8}
\cline{2-8}
\cline{3-8}
\cline{4-8}
\multirow{3}{*}{\makecell[l]{Male\\names}} & \multirow{3}{*}{\makecell[l]{Female\\names}} & \multirow{3}{*}{Career} & \multirow{3}{*}{Family} & all &  2.78**    &  -0.23\hphantom{**}   &    2.16**   \\
         &            &                     &         & female &  0.45\hphantom{**}    &   -0.53\hphantom{**}   &    0.98\hphantom{**}     \\
         &            &                     &         & male &   2.76**   &  -0.60\hphantom{**} &   2.56**     \\

\cline{1-8}
\cline{2-8}
\cline{3-8}
\cline{4-8}
\multirow{12}{*}{\makecell[l]{Male\\nouns \&\\pronouns }} & \multirow{12}{*}{\makecell[l]{Female\\nouns \&\\pronouns }} & \multirow{3}{*}{Career} & \multirow{3}{*}{Family} & all &     0.64\hphantom{**} &     0.15\hphantom{**} &        0.57\hphantom{**} \\
         &            &                     &         & female &    -0.22\hphantom{**} &    -0.34\hphantom{**} &        0.04\hphantom{**} \\
         &            &                     &         & male &     1.91\hphantom{**} &     0.35\hphantom{**} &        0.38\hphantom{**} \\
\cline{3-8}
\cline{4-8}
         &            & \multirow{3}{*}{MatSci} & \multirow{3}{*}{Arts} & all &     0.50\hphantom{**} &    -0.45\hphantom{**} &        0.94\hphantom{**} \\
         &            &                     &         & female &     0.06\hphantom{**} &  -1.26** &       1.29*\hphantom{*} \\
         &            &                     &         & male &     0.39\hphantom{**} &    -0.09\hphantom{**} &        0.37\hphantom{**} \\
\cline{3-8}
\cline{4-8}
\cline{3-8}
\cline{4-8}
         &            & \multirow{3}{*}{\bfseries\makecell[l]{Intelligence*\\0.18}} & \multirow{3}{*}{Appearance} & all &     0.29\hphantom{**} &    -0.13\hphantom{**} &        0.39\hphantom{**} \\
         &            &                     &         & female &    -0.08\hphantom{**} &    -0.09\hphantom{**} &        0.06\hphantom{**} \\
         &            &                     &         & male &     0.42\hphantom{**} &    -0.31\hphantom{**} &       0.71*\hphantom{*} \\
\cline{3-8}
\cline{4-8}
         &            & \multirow{3}{*}{\bfseries\makecell[l]{Strength**\\0.32}} & \multirow{3}{*}{Weakness} & all &   1.71** &     0.40\hphantom{**} &      1.16** \\
         &            &                     &         & female &    0.70*\hphantom{*} &     0.24\hphantom{**} &        0.32\hphantom{**} \\
         &            &                     &         & male &   1.24** &     0.36\hphantom{**} &      1.02** \\
\hline
\end{tabular}
}
\end{table}

Finally, we find a positive and slightly significant WEAT score (0.71) comparing the Intelligence and Appearance words in male lyrics. Although the single category scores are not significant by themselves, their opposite sign shows a signal for male artists to associate Intelligence words with Male terms and Weakness words with Female terms. This is different in female lyrics where the corresponding SC-WEAT scores are close to zero indicating no bias in this regard and translates into a slightly significant result for the SWEAT score for Intelligence (0.18).

To summarise, this analysis shows that songs of male solo artists contain more and often stronger gender biases than those of female solo artists, which are closer to gender neutrality. The only exception of this observation is a bias in female lyrics relating females more closely to art terms.
Male artist songs emphasize men as stronger and focused on career at the expense of women depicted as less strong and closer to the family-related terms. 

\section{Discussion}
This section discusses our findings in relation to the research questions and findings of previous works, some potential shortcomings in our study design, and potential paths for future research.  

\textbf{Sexism (RQ1)}: The first research question investigates the presence of sexist content in song lyrics and whether there is any difference between artist gender, type, and musical genre. We found that almost 25\% of song lyrics express sexist content, but this share is not uniformly distributed across artist gender and type. Male solo artists have the highest share of sexist lyrics in all the three subsets of the WASABI dataset (all songs, songs in the Billboards charts, and songs in top 10 position). 
Interestingly, this share is much lower for groups composed of only male members. This difference is worth mentioning because it might indicate a deeper divergence between songs performed by male solo and group artists that, to our knowledge, has not been reported in the literature previously. 

Another observation is that the relative number of lyrics containing sexist passages is higher in Billboard charts than in the whole WASABI dataset independently from the gender and type of the artist, a trend that is stronger going forward in time. Other works found similar observations, even though using different definitions or specific aspects related to sexual content. 
For example, some studies report an increase in sexual content and objectification of women in popular songs during the last five decades~\cite{smiler2017want, Madanikia2014themes}. In a study limited to the year 2009, the top 10 song charts were shown to be more likely to contain sexual content if compared to songs from the same album by the same artist that did not enter the top 10~\cite{hobbs2011songs}. According to previous studies, we also found an higher fraction of songs containing sexist content among Hip hop and R\&B and soul songs~\cite{Hart2020Linguistic, flynn2016objectification}, albeit other genres are not exempted from this trend~\cite{Neff2014SexismAM}. Hip hop and R\&B and soul were also found to display an increasing trend in associating competence more frequently to men than women~\cite{Boghrati2022Quantifying}.

Although we obtained aggregated results in line with findings from previous works, we emphasize that these results rely on a sexism classifier trained on out-of-domain data, which might have led to poor generalization when applied to song lyrics. However, the classifier was trained on a combination of diverse corpora of sexist texts that enforce its ability to generalize on out-of-domain data~\cite{samory2021call}. We have also validated our model on a dataset of sexist lyrics showing good performances and robust results across different classification thresholds. We believe that this exploratory data analysis can encourage further works aimed at identifying sexism in song lyrics, which may need to pass through extensive manual labeling in order to train a classifier on the specific domain. 

\textbf{Language bias (RQ2)}: 
The second research question focuses on the differences of language bias in male and female solo artist song lyrics. 
Our results extend the ones obtained in~\cite{barman2019decoding}, where all the WEAT scores are positive in magnitude. We enrich this work in two different ways. First, we measure language biases separately in the subsets of male and female solo artist songs, obtaining all positive WEAT scores as well. Some of these associations (e.g., male with career and female with family) align to the ones observed in a large-scale crawl of the Internet~\cite{caliskan2017semantics}, indicating how music reflects societal biases. Second, we enrich the analysis through other association tests, which measure associations against a single target set (SC-WEAT) and between the two corpora (SWEAT). Our results show that the biases affecting these two corpora are different. In particular, male lyrics contain larger gender biases. For instance, Female proper names are more associated to Family than Career words, while Career words are closer to Male than Female names. At the same time, Strength words are more associated to Male than Female terms. The female corpus does not contain these two types of biases. Similarly, other works have found song lyrics of male artists to contain stronger gender biases than female artists, in particular depicting men as more competent than women~\cite{Boghrati2022Quantifying}.

In both corpora, we find that Strength words are closer to Male than Female terms but this association is again stronger in male lyrics. A bias that is only present in female lyrics is that female solo artists use more frequently Female terms closer to Art than Mathematics and Science words. Interestingly, in addition to a weak association of Intelligence words with Male terms in the male corpus, there is no significant association between Appearance words and gendered terms. This might be related to the observation that objectification does not target uniquely women, but it is also fairly common for men even though not present to the same extent~\cite{flynn2016objectification}. Besides these gender biases, our word embeddings can capture expected associations like Pleasant with Flower words and Unpleasant with Insects or Weapons that do not depend on the gender of the artist.

Our results lack of an analysis of gender biases across time and genres as done in Boghrati and Berger~\cite{Boghrati2022Quantifying}, and a comparison between solo artists and groups songs. 
However, these subsets of the dataset would have reduced the training data used to learn word embeddings, thus compromising their quality. 

Other limitations concern the dataset. We can not claim any generalizability of our conclusions because there is no guarantee that WASABI contains a representative sample of English songs. Moreover, Billboard charts add an additional US bias in the selection of popular songs. The choice of the variables used to stratify the results is another limitation. It is worth noticing that we split artists into males and females taking into account only the performer of the song, thus ignoring the gender of the songwriter. In addition, we can consider only two
genders, thus neglecting other non-binary gender identities, as only binary gender data is available in the WASABI database. It would be interesting to address these gaps in future works, but this will require
additional efforts for accurate data collection.
Despite these limitations, the size of WASABI database, together with its open access, makes it a relevant study object by itself, and our contribution may be valuable for future users of this resource. 

\section{Conclusion}

In this work, we have exploited the WASABI database to describe how sexist content in music varies during five decades, from 1960 to 2010, and to what extent word embeddings learned from the song lyrics corpus contain language biases. The former analysis, stratified by artist gender and type, shows that popular song lyrics of male solo artists become more sexist over the years, while this behavior is less noticeable for the other categories of artists. The genres (among the ones analyzed) that have the highest fraction of songs containing sexist content are Hip hop and R\&B and soul, independently from the gender of the artist. Regarding language biases, we find the lyrics of male solo artists to contain more gender bias than those of female solo artists. This is true for instance for the stereotype depicting men as stronger and focused on success, and women being closer to family. The former bias is present in female solo artist songs as well, even though the association is weaker than for male solo artists. Our results show different ways to extract meaningful metrics about language usage and bias in song lyrics, as well as how to analyse such an important and heavily consumed expression of popular culture that influences how listeners see the world and reflects how artists perceive it.

\clearpage
\newpage

\section*{Declarations}

\section*{Availability of data and materials}
Instructions to download the dataset analysed during the current study and code produced for the analysis are available in the author’s GitHub repository: \url{https://github.com/Loreb92/sexism_and_bias_in_song_lyrics}.

\section*{Competing interests}
  The authors declare that they have no competing interests.

\section*{Funding}
The authors acknowledge support from Intesa Sanpaolo Innovation Center. The funder had no role in study design, data collection and analysis, decision to publish, or preparation of the manuscript.

\section*{Author's contributions}
LB run the analysis. AK conceptualized the problem and collected the data. All authors contributed to interpret the results and to the writing of the manuscript.
All authors read and approved the final manuscript.

 \bibliographystyle{elsarticle-num} 
 \bibliography{cas-refs}

\clearpage

\newpage
\appendix

\section{Genre Popularity}\label{appendix:descriptive_stats}

\begin{figure*}[!tbh]
    \centering
    \includegraphics[width=.95\textwidth]{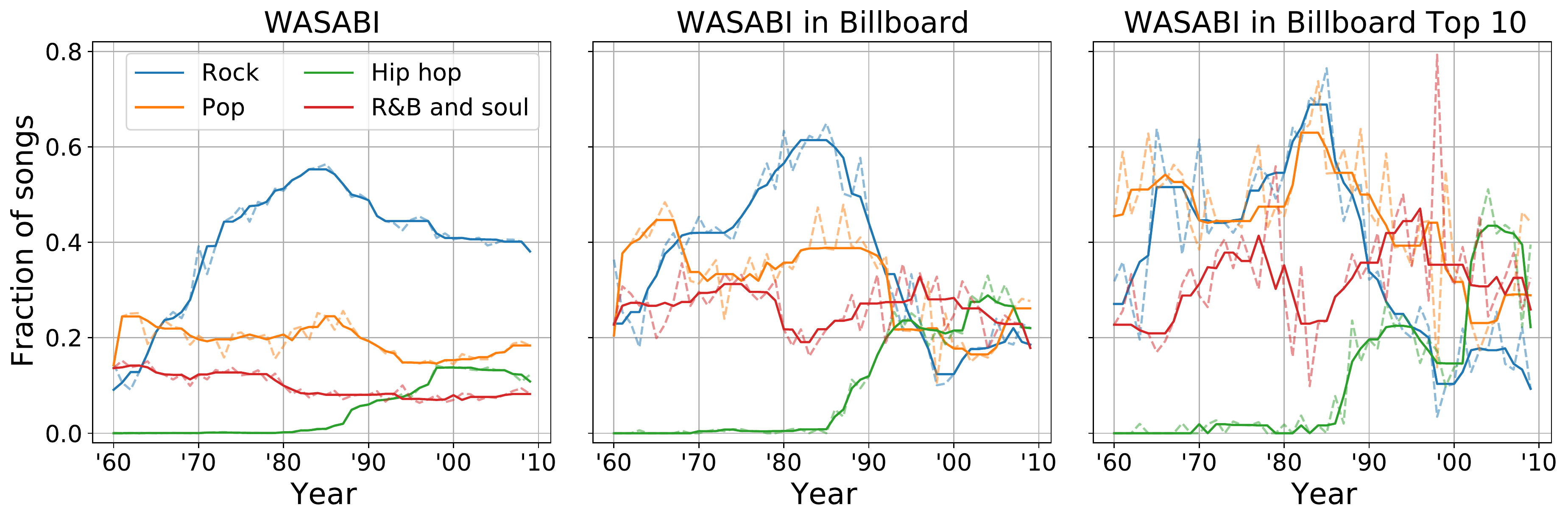}
    \caption{Fraction of songs in our data sets for four genres: Pop, Rock, Hip hop, and R\&B and soul. }
    \label{fig:frac_songs_across_datasets_time_genre}
\end{figure*}

We show here in Figure~\ref{fig:frac_songs_across_datasets_time_genre} the fraction of Rock, Pop, Hip hop and R\&B and soul songs across time for the three subsets of the WASABI dataset.
The popularity of Rock peaked between 1970 and 1990 and then decreased significantly going forward in time, although it is the most frequent genre in WASABI. Pop songs display a similar decrease during the same period of time in Billboard charts. On the other hand, R\&B and soul represents more than 20\% of songs on Billboard during the whole time span, while Hip hop starts to flourish during the mid-80s.

\section{Duplicate lyrics detection}\label{appendix:duplicate_detection}

The WASABI database was created by collecting the discography of all the artists, including the text of song lyrics when available. This latter information was collected from LyricWiki, an online database of lyrics and encyclopedia curated by users~\cite{meseguer2017wasabi}. During this process, the same song lyrics may be collected multiple times (e.g., the same song is present in several albums of the same artist) and it is ideal to reduce as much as possible the presence of duplicates in the dataset. To do that, we used techniques of approximate string matching.

Given a set of song lyrics, we represented each lyrics as the set of 3-grams it is composed of, following a bag-of-words approach. Then, we consider two lyrics to be duplicates if the Jaccard index of their 3-grams representation is higher than $0.80$. This threshold allowed us to detect groups of songs whose lyrics are identical up to slight variations. We then consider the song with the earliest publication date as the original song and the other songs of the group as its duplicate. 

In case two duplicate songs were performed by two distinct artists, we refer to the song published later as cover song. We removed from the dataset only the duplicate songs but we kept cover versions. We identified \num{82531} duplicate songs and \num{7524} cover songs.

\section{Matching songs from different datasets}\label{appendix:matching_datasets}

Due to the lack of standardized metadata of artist names and song titles, it is not trivial to match entries corresponding to the same artist or song from two datasets. In our case, we needed to match the entries of the WASABI dataset with the entries of the Billboard charts and the sexist lyrics dataset. We used hand crafted logics of approximate string matching to first match the artist name, and then the title of the song. We used Python script from the Million Song Dataset repository\footnote{\url{https://github.com/tbertinmahieux/MSongsDB/blob/master/NameNormalizer/normalizer.py}}, which we only modified slightly. 

\section{Sexism classifier on song lyrics}\label{appendix:scores_sexism_classif}

\begin{table}[t!]
\caption{Performance of the sexism classifier on the external dataset for three classification thresholds and $N_B=1$. Metrics for both classes (Sexist and Not sexist) and the corresponding macro average are shown. Right column shows the performance of a naive baseline that always predicts the sexist class.}
\label{tab:basic_stats_sexism_classifier}
\centering
 \begin{tabular}{@{}l|l|r|r|r|r@{}}   
 Metric& Class & \multicolumn{3}{c|}{Classif. threshold} & Baseline\\
  & &  0.50 & 0.725 & 0.90 & \\
 \hline 
  \hline 
\multirow{3}{*}{Precision} & Sexist & 0.62 & 0.68 & 0.78 & 0.41 \\
& Non-sexist & 0.78 & 0.79 & 0.71 & 0.00 \\
& Macro avg.  & 0.70 & 0.73 & 0.74 & 0.20 \\
\hline
\multirow{3}{*}{Recall} & Sexist & 0.70 & 0.69 & 0.45 & 1.00 \\
& Non-sexist & 0.71 & 0.78 & 0.91 & 0.00 \\
& Macro avg. &  0.70 & 0.73 & 0.68 & 0.50\\
\hline
\multirow{3}{*}{F1-score} & Sexist & 0.66 & 0.68 & 0.57 & 0.58 \\
& Non-sexist & 0.74 & 0.78 & 0.80 & 0.00\\
 & Macro avg. & 0.70 & 0.73 & 0.69 & 0.29
    \end{tabular}    
\end{table}

We trained the classifier using the scripts provided by the authors of~\cite{samory2021call} and stored in the official repository of the paper.\footnote{\url{https://github.com/gesiscss/theory-driven-sexism-detection}} The pre-trained model is the base version of the uncased BERT\footnote{\url{https://tfhub.dev/google/bert\_uncased\_L-12\_H-768\_A-12/1}}, fine tuned for 3 epochs with batch size equal to 32, learning rate at $2\times10^{-5}$ and 10\% of warmup steps. 

Since the classifier we used to detect sexist lyrics was trained on short texts, we adapted the representation of the input to classify song lyrics. In detail, lyrics were divided into batches composed of groups of four lines, each of them sharing two lines with the previous and following group. A batch is classified as sexist if the model outputs a probability higher than a certain threshold. Then, lyrics are classified as sexist whenever the model predicts that at least $N_B$ batches contain sexist content. 
In the main text, we considered $N_B = 1$, meaning that one batch of lines labeled as sexist is enough to propagate the label to the whole song. The optimal classification threshold is chosen as the threshold maximizing the F1-score on the external dataset~\cite{sexist_dataset} of sexist songs in order to balance both precision and recall. This corresponds to a threshold equal to 0.725.

Table~\ref{tab:basic_stats_sexism_classifier} shows the classification scores on the external dataset and the performance of a naive baseline model that predicts all lyrics as sexist. Besides the optimal classification threshold, we also considered the results for classification thresholds at $0.50$ and $0.90$, where the latter favors precision on the sexist class over recall, and minimises false positives. Indeed, with the $0.90$ threshold the precision for the sexist class reaches 0.78 while the macro F1-score is 0.69 and the recall drops to 0.45. At the same time, the performance on the non-sexist class reaches a recall of 0.91 with a precision of 0.71. 

\begin{figure}
    \centering
    \includegraphics[width=.85\textwidth]{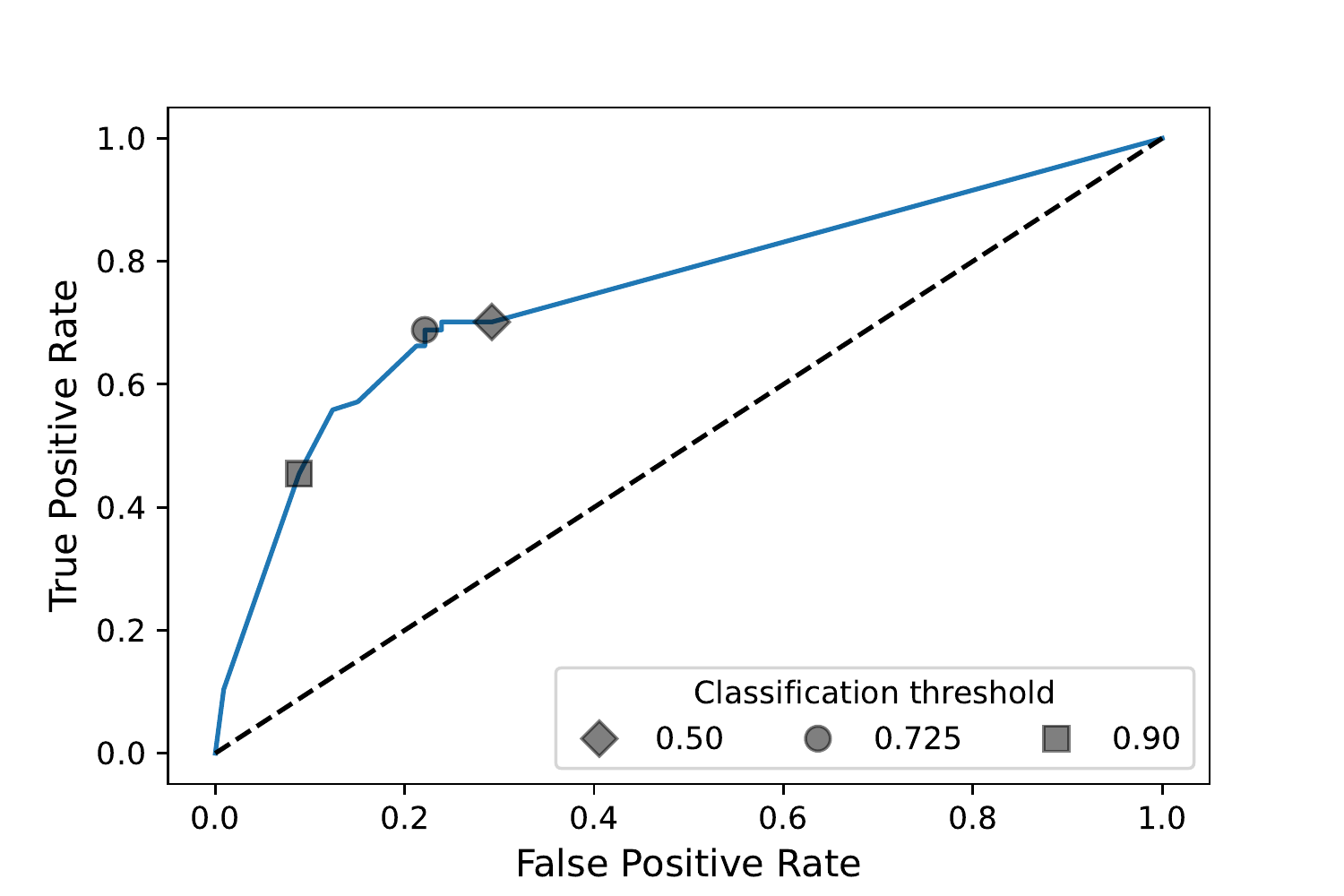}
    \caption{ROC curve of the sexism classifier for $N_B=1$. Markers indicate the points corresponding to the three classification thresholds discussed.}
    \label{fig:roc_auc_sexism_classif}
\end{figure}

The corresponding ROC curve is shown in Figure~\ref{fig:roc_auc_sexism_classif}.  We also note that the F1-score on the sexist class is stable for other choice of $N_B$ ($N_B = 2$: F1-score at 0.69 for an optimal threshold of 0.575; $N_B = 3$: F1-score at 0.65 for an optimal threshold of 0.525).

\begin{figure}
    \centering
    \includegraphics[width=.95\textwidth]{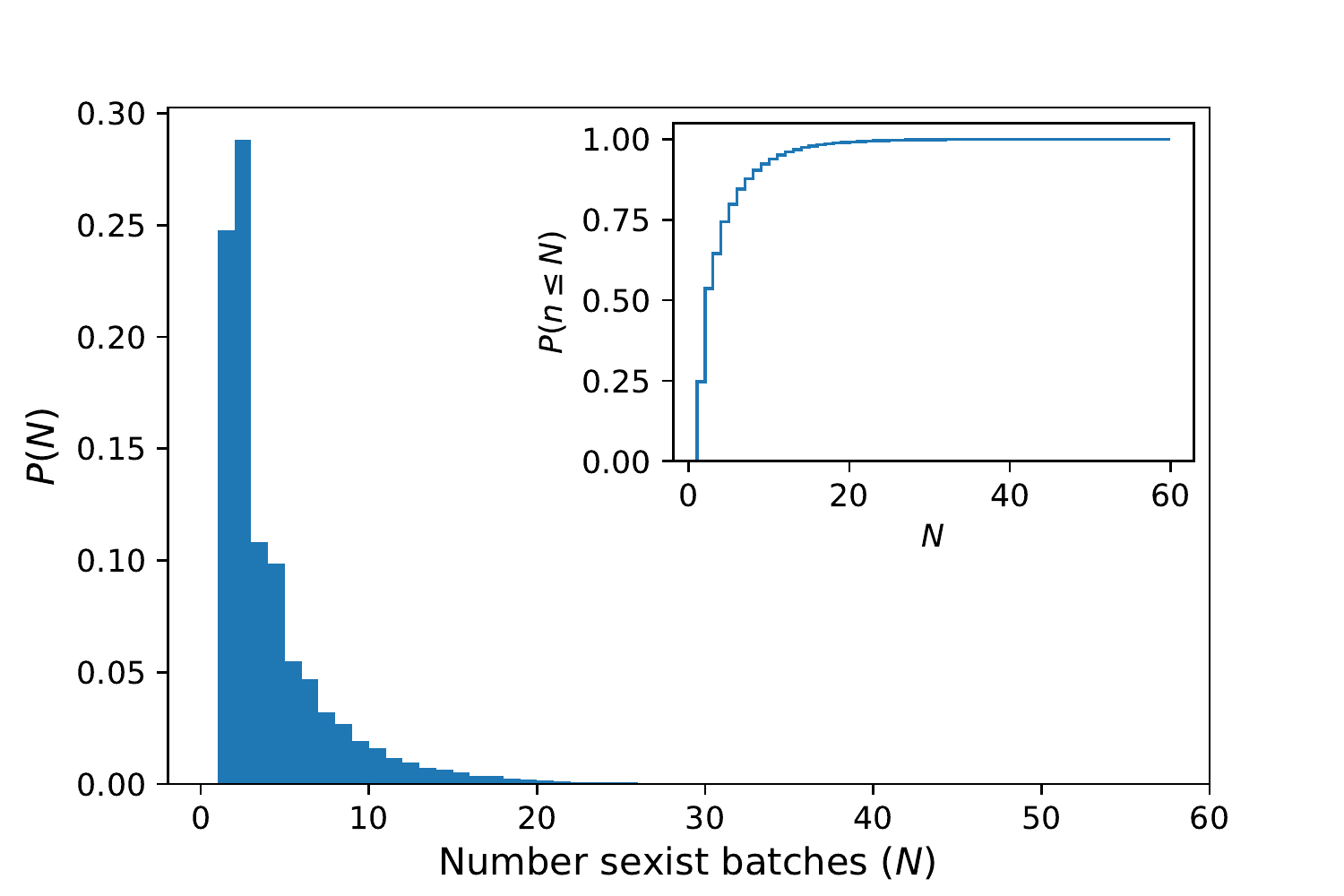}
    \caption{Distribution of the number $N$ of batches per song classified as containing sexist content (ignoring songs with $N=0$) with classification threshold at $0.725$. The maximum number of sexist batches in a song is 60. The cumulative distribution is shown in the inset.}
    \label{fig:distribution_sexist_flags}
\end{figure}

Now we come back to discuss the results of the sexism classifier on the WASABI dataset. Figure~\ref{fig:distribution_sexist_flags} shows the distribution of sexist batches for songs classified as sexist according to the optimal classification threshold and $N_B = 1$. Almost half of the lyrics have more than 2 sexist batches, while 25\% and 29\% of song lyrics have 1 and 2 sexist batches respectively.

To verify the independence of the conclusions we extracted from the optimal classification threshold, we report the fraction of sexist lyrics for all the three thresholds on Table~\ref{tab:basic_stats_dataset_sexism}. Here, we can observe for all the thresholds the same trend showing that (i) male solo artists performs more sexist songs compared to the other groups, and (ii) Billboard charts contain a higher fraction of sexist songs than the whole WASABI dataset, regardless of the gender and type of the performer. Table~\ref{tab:basic_stats_dataset_sexism_n_flags} shows that these considerations hold for $N_B = 2$ and $N_B = 3$ as well.

\begin{table}[ht!]
\caption{Percentage of sexist songs identified for each artist type and gender. Results for three different classification thresholds (0.50, 0.725 and 0.90) and $N_B=1$ are reported, with results for the optimal threshold in bold. Percentages correspond to the fraction of sexist lyrics within the artist type and gender.}
\label{tab:basic_stats_dataset_sexism}
\centering
    \begin{tabular}{@{}l|l|l||r|r|r@{}} 
       \makecell{Artist\\Type} & Gender & \makecell{Classif.\\ thresh.} & Songs & Billboard & \makecell{Billboard\\(top 10)} \\
        \hline
        \hline
        \multirow{6}*{\makecell{Solo\\Artist}} & \multirow{3}*{\makecell{Male}} & 0.50 & \makecell[r]{36.7\%} & \makecell[r]{50.0\%} & \makecell[r]{54.1\%}  \\
        & & \bfseries 0.725 & \bfseries \makecell[r]{30.0\%} & \bfseries \makecell[r]{43.4\%} & \bfseries \makecell[r]{48.4\%} \\
        & & 0.90 & \makecell[r]{11.9\%} & \makecell[r]{22.0\%} & \makecell[r]{28.0\%} \\ \cline{2-6}
        & \multirow{3}*{\makecell{Female}} & 0.50 & \makecell[r]{25.0\%} &  \makecell[r]{35.7\%} & \makecell[r]{36.9\%} \\
        & & \bfseries0.725 &\bfseries\makecell[r]{19.6\%} & \bfseries\makecell[r]{29.8\%} & \bfseries\makecell[r]{30.9\%} \\
        & & 0.90 &\makecell[r]{7.8\%} & \makecell[r]{14.6\%} & \makecell[r]{16.0\%} \\ \hline
        \multirow{9}*{Group} & \multirow{3}*{\makecell[l]{Male \\only}} & 0.50 
        & \makecell[r]{23.2\%} & \makecell[r]{39.1\%} & \makecell[r]{39.7\%} \\
         & & \bfseries  0.725 &\bfseries \makecell[r]{18.0\%} & \bfseries \makecell[r]{32.0\%} & \bfseries \makecell[r]{33.2\%} \\
         & & 0.90 &\makecell[r]{6.3\%} & \makecell[r]{13.6\%} & \makecell[r]{14.8\%} \\ \cline{2-6}
        & \multirow{3}*{\makecell[l]{Female \\only}} & 0.50 & \makecell[r]{25.2\%} & \makecell[r]{42.5\%} & \makecell[r]{50.0\%} \\
         & &  \bfseries 0.725 &\bfseries \makecell[r]{20.0\%} & \bfseries \makecell[r]{34.4\%} & \bfseries \makecell[r]{42.9\%} \\
         & & 0.90 &\makecell[r]{8.3\%} & \makecell[r]{14.6\%} & \makecell[r]{21.4\%} \\ \cline{2-6}
        & \multirow{3}*{\makecell{Mixed}} & 0.50  & \makecell[r]{21.5\%} & \makecell[r]{32.8\%} & \makecell[r]{37.2\%} \\
         & & \bfseries 0.725 & \bfseries \makecell[r]{16.2\%} & \bfseries \makecell[r]{26.4\%} & \bfseries \makecell[r]{30.1\%} \\ 
          & & 0.90 & \makecell[r]{5.2\%} & \makecell[r]{13.3\%} & \makecell[r]{16.0\%} \\ 
    \end{tabular}
\end{table}

\begin{table}[ht!]
\caption{Percentage of sexist songs identified for each artist type and gender. We report how these percentages change for different values of $N_B$ (i.e. the minimum number of 4 line batches classified as sexist to consider the whole song to contain sexist content). The classification threshold $=0.725$ and values in bold are used in the main text. Percentages correspond to the fraction of sexist lyrics within the artist type and gender. }
\label{tab:basic_stats_dataset_sexism_n_flags}
\centering
    \begin{tabular}{@{}l|l|l||r|r|r@{}} 
       \makecell{Artist\\Type} & Gender & \makecell{$N_B$} & Songs & Billboard & \makecell{Billboard\\(top 10)} \\
        \hline
        \hline
        \multirow{6}*{\makecell{Solo\\Artist}} & \multirow{3}*{\makecell{Male}} & \bfseries \makecell[c]{1} & \bfseries \makecell[r]{30.0\%} & \bfseries \makecell[r]{43.4\%} & \bfseries \makecell[r]{48.4\%} \\
        & & \makecell[c]{2} &  \makecell[r]{23.1\%} &  \makecell[r]{36.7\%} &  \makecell[r]{42.1\%} \\
        & & \makecell[c]{3} & \makecell[r]{14.9\%} & \makecell[r]{25.5\%} & \makecell[r]{32.8\%} \\ \cline{2-6}
        & \multirow{3}*{\makecell{Female}} & \bfseries \makecell[c]{1} & \bfseries\makecell[r]{19.6\%} & \bfseries\makecell[r]{29.8\%} & \bfseries\makecell[r]{30.9\%} \\
        & & \makecell[c]{2} & \makecell[r]{14.7\%} & \makecell[r]{23.8\%} & \makecell[r]{25.5\%} \\
        & & \makecell[c]{3} &\makecell[r]{8.6\%} & \makecell[r]{14.9\%} & \makecell[r]{16.9\%} \\ \hline
        \multirow{9}*{Group} & \multirow{3}*{\makecell[l]{Male \\only}} & \bfseries \makecell[c]{1} 
        & \bfseries \makecell[r]{18.0\%} & \bfseries \makecell[r]{32.0\%} & \bfseries \makecell[r]{33.2\%} \\
         & &  \makecell[c]{2}  & \makecell[r]{12.9\%} &  \makecell[r]{24.6\%} &  \makecell[r]{26.2\%} \\
         & & \makecell[c]{3} &\makecell[r]{7.4\%} & \makecell[r]{14.2\%} & \makecell[r]{16.7\%} \\ \cline{2-6}
        & \multirow{3}*{\makecell[l]{Female \\only}} & \bfseries \makecell[c]{1} & \bfseries \makecell[r]{20.0\%} & \bfseries \makecell[r]{34.4\%} & \bfseries \makecell[r]{42.9\%} \\
         & &  \makecell[c]{2}  & \makecell[r]{15.0\%} &  \makecell[r]{26.4\%} &  \makecell[r]{37.5\%} \\
         & & \makecell[c]{3} &\makecell[r]{8.4\%} & \makecell[r]{14.2\%} & \makecell[r]{17.9\%} \\ \cline{2-6}
        & \multirow{3}*{\makecell{Mixed}} & \bfseries \makecell[c]{1}  & \bfseries \makecell[r]{16.2\%} & \bfseries \makecell[r]{26.4\%} & \bfseries \makecell[r]{30.1\%} \\
         & & \makecell[c]{2} &  \makecell[r]{11.2\%} &  \makecell[r]{20.8\%} &  \makecell[r]{25.6\%} \\ 
          & & \makecell[c]{3} & \makecell[r]{6.3\%} & \makecell[r]{13.8\%} & \makecell[r]{17.9\%} \\ 
    \end{tabular}
\end{table}

\begin{table*}[!tb]

\caption{Word sets used for the word embedding association tests. The right column reports the references from which the sets were borrowed.}
\label{tab:words_used}
\resizebox{\textwidth}{!}{
\begin{tabular}{@{}p{2.2cm}p{18.6cm}p{0.7cm}@{}}
\toprule
Word set name & Words & Ref. \\
\toprule

Pleasant & family, honest, gift, wonderful, vacation, miracle, loyal, pleasure, gentle, rainbow, love, peace, lucky, honor, freedom, happy, health, friend, laughter, cheer, joy, heaven, diploma, paradise, diamond, caress, sunrise & \cite{caliskan2017semantics} \\[0.5cm]
Unpleasant & cancer, agony, stink, sickness, poverty, accident, failure, rotten, hatred, terrible, disaster, tragedy, grief, jail, abuse, awful, prison, ugly, nasty, murder, bomb, poison, evil, crash, death, war, kill & \cite{caliskan2017semantics} \\[0.5cm]
\hline
Career & corporation, professional, career, office, business & \cite{caliskan2017semantics, chaloner2019measuring} \\[0.1cm]
Family & family, marriage, wedding, children, home & \cite{caliskan2017semantics, chaloner2019measuring} \\[0.1cm]
\hline
Female attributes & girl, hers, her, aunt, daughter, sister, female, mother, she, grandmother, woman & \cite{caliskan2017semantics, chaloner2019measuring} \\[0.1cm]
Male attributes & brother, grandfather, his, son, father, man, male, uncle, he, him, boy & \cite{caliskan2017semantics, chaloner2019measuring}  \\[0.1cm]
\hline
Female names & kim, rose, mary, eve, kelly, jane, lisa, juliet, jean, annie, trina, sarah, sally, betty, lucy, taylor, bonnie, marie, jenny, dolly, julia, anna, jill, angie & *\\[0.5cm]
Male names & john, jack, joe, johnny, james, david, paul, billy, jimmy, simon, mark, romeo, bill, peter, bob, lee, jim, bobby, tom, jackson, sam, michael, charlie, adam & * \\[0.5cm]
\hline
Flowers & lilac, bluebell, violet, crocus, buttercup, iris, rose, tulip, daisy, marigold, daffodil, orchid, carnation, magnolia, lily, poppy, clover & \cite{caliskan2017semantics} \\[0.5cm]
Insects & cockroach, maggot, locust, roach, centipede, caterpillar, weevil, beetle, flea, dragonfly, mosquito, ant, cricket, moth, spider, bee, fly & \cite{caliskan2017semantics} \\[0.5cm]

\hline
Musical instruments & banjo, mandolin, trombone, cello, fiddle, tuba, harmonica, harp, violin, piano, trumpet, clarinet, oboe, guitar, lute, saxophone, horn, bongo, flute, bell, viola, drum & \cite{caliskan2017semantics} \\[0.5cm]
Weapons & harpoon, mace, hatchet, grenade, missile, spear, axe, rifle, cannon, dagger, pistol, shotgun, dynamite, tank, blade, sword, arrow, whip, bomb, knife, club, gun & \cite{caliskan2017semantics} \\ [0.5cm]
\hline
MatSci  & nasa, addition, einstein, technology, experiment, math, chemistry, science & \cite{caliskan2017semantics, chaloner2019measuring}  \\[0.1cm]
Arts words & poetry, novel, symphony, art, dance, shakespeare, sculpture, drama & \cite{caliskan2017semantics, chaloner2019measuring} \\[0.1cm]
\hline
Intelligence  & brilliant, logical, apt, smart, thoughtful, wise, precocious, genius, intelligent, shrewd, clever & \cite{chaloner2019measuring} \\[0.1cm]
Appearance  & gorgeous, slim, healthy, handsome, ugly, fat, thin, weak, beautiful, pretty, strong & \cite{chaloner2019measuring} \\[0.1cm]
\hline

\hline

Strength  & triumph, confident, potent, loud, winner, shout, succeed, strong, bold, leader, dynamic, command, power & \cite{chaloner2019measuring} \\[0.1cm]
Weakness  & timid, withdraw, yield, failure, fragile, weakness, loser, shy, surrender, weak, afraid, follow, lose &  \cite{chaloner2019measuring}\\
\bottomrule
\multicolumn{3}{l}{* Male and female names were chosen from the most frequent names found in the solo artists songs of the WASABI dataset.}\\

\end{tabular}
}

\end{table*}

\section{Word embedding association tests}\label{appendix:word_sets}

Table~\ref{tab:words_used} shows the lists of words corresponding to each set of target and attribute words, and we report the references from which each word set was borrowed on the right column. We applied slight modifications to some of those sets to take into account rare or missing words. First, we removed words if they occur less than 5 times in one of the three corpora. Then, whenever pairs of attribute (target) sets contain a different amount of words, we remove the least frequent words from the larger until the two sets have the same size. In doing this, we define the frequency of a word as the minimum of the frequencies among the three corpora under analysis. We used a different procedure to select proper names. We downloaded the yearly counts of names of newborn babies from  1879 until 2020\footnote{\url{https://www.ssa.gov/oact/babynames/limits.html} (accessed 12/01/2022)}, and searched for them in song lyrics. The male and female name word sets are thus composed of the most frequent names in the lyrics corpora.

{\textbf{Learning word vectors}:}
We used the Gensim implementation of Word2Vec~\cite{rehurek2011gensim} to learn word embeddings with window length 5, embedding dimension 300, and 40 training steps. Words occurring less than 5 times were discarded. 

{\textbf{Detailed results on Pleasant vs. Unpleasant words}:} Results are shown in Table~\ref{tab:results_weat}. The coupled association measured by the WEAT returns positive and statistically significant scores for all the three corpora and both pairs of target sets, namely Flowers/Insects and Musical instruments/Weapons. The single category associations (SC-WEAT) of the former are significant as well, indicating that, if kept separately, Flower words are more associated to Pleasant than Unpleasant words and vice versa for Insect words. For the latter, the only female corpus shows a significant single category association between Musical instrument and Pleasant words. This reflects in the comparison between the male and female corpus through the SWEAT that returns a negative and statistically significant score, i.e., Musical instrument words are closer to Unpleasant than Pleasant words in the male corpus while being closer to Pleasant than Unpleasant words in the female corpus. The fact that the learned word embeddings encode these associations makes us confident of the quality of the learned word vectors.





\end{document}